\documentclass[12pt]{article}
\usepackage{amsmath}
\usepackage{amssymb}
\date{}
\begin{document}

\title{Convective Equations and a Generalized Cole-Hopf Transformation}

\author{Mayer Humi\\
Department of Mathematical Sciences,\\
Worcester Polytechnic Institute,\\
100 Institute Road,\\
Worcester, MA  0l609}

\maketitle

\begin{abstract}

Differential equations with convective terms such as the Burger's equation
appear in many applications and have been the subject of intense research. 
In this paper we use a generalized form of Cole-Hopf transformation to relate 
the solutions of some of these nonlinear equations to the solutions of linear 
equations. In particular we consider generalized forms of Burger's equation
and second order nonlinear ordinary differential equations with 
convective terms which can represent steady state one-dimensional convection.
\end{abstract}

\thispagestyle{empty}

\newpage

\section{Background}

Equations with convective terms appear in various applications
in applied mathematics and theoretical physics. In particular
Burger's equation [1]
\begin{equation}
\label{1.1}
\frac{\partial\psi}{\partial t}+\psi\frac{\partial\psi}{\partial x}=
\nu\frac{\partial^2\psi}{\partial x^2},\,\,\,\, \nu=constant
\end{equation}
which serves as a prototype for turbulence modeling, gas dynamics and 
traffic flow contains such a term. However it was found that this equation 
can be linearized and reduced to the linear Heat equation
\begin{equation}
\label{1.2}
\frac{\partial\psi}{\partial t}=k\frac{\partial^2\psi}{\partial x^2}
\end{equation}
by Cole-Hopf transformation [2,3]
\begin{equation}
\label{1.3}
\psi=\frac{\frac{\partial\phi}{\partial x}}{\phi}.
\end{equation}
Since this discovery many attempts were made in the literature to 
generalize this result to larger class of equations[4-10] or to relate other 
nonlinear equations to this equation[8].

Recently however, we introduced [11] a generalization of Cole-Hopf 
transformation and used it to linearize various nonlinear ordinary 
differential equations e.g Duffing equation. We showed also that many of the 
special functions of mathematical physics are exact solutions for a class 
of nonlinear equations.

In this paper we apply in Sec. 2 the same generalization of (\ref{1.3})
to linearize a class of Burger's equations with additional quadratic 
nonlinear terms. Next in Sec. 3 we consider second nonlinear ordinary 
differential equations(ODEs) with convective term and derive conditions under 
which they can be linearized by the same transformation. These type of
equations represent steady state convection in one dimension [12,13].
We end up in Sec. 4 with some conclusions.

\setcounter{equation}{0} 
\section{Generalized Burger Equation}
In this section we consider equations of the form
\begin{equation}
\label{4.1}
\frac{\partial\psi}{\partial t}-M(x)\frac{\partial^2\psi}{\partial x^2} =
H(x)\psi\frac{\partial\psi}{\partial x}+V(x)\psi+W(x)\psi^2
\end{equation} 
where $\psi=\psi(x,t)$.
We shall say that the solutions of this equation are related
to the solution of the linear equation
\begin{equation}
\label{4.2}
\frac{\partial\phi}{\partial t}-M(x)\frac{\partial^2\phi}{\partial x^2}=0
\end{equation}
if we can find functions $P(x)$, $Q(x)$ so that
\begin{equation}
\label{4.3}
\psi=P(x)+Q(x)\frac{\frac{\partial\phi}{\partial x}}{\phi}
\end{equation}

To classify those equations of the form (\ref{4.1}) which can be paired 
to the linear equation (\ref{4.2}) we substitute (\ref{4.3}) in (\ref{4.1}).
After some algebra we find that the following equation has to be satisfied
\begin{equation}
\label{4.4}
a_3\left(\frac{\frac{\partial\phi}{\partial x}}{\phi}\right)^{3}+
a_2\left(\frac{\frac{\partial\phi}{\partial x}}{\phi}\right)^{2}+
a_{21}\left(\frac{\frac{\partial\phi}{\partial x}}{\phi^2}\right)+
a_1\frac{\frac{\partial\phi}{\partial x}}{\phi}+
\frac{a_{11}}{\phi(x)}+
a_0=0,
\end{equation}
where
\begin{equation}
\label{4.5}
a_{21}=-Q(x)\left[(Q(x)H(x)-3M(x))\frac{\partial^2\phi}{\partial x^2}+
\frac{\partial\phi}{\partial t}\right],
\end{equation}
\begin{equation}
\label{4.6}
a_{11}=-(Q(x)H(x)P(x)+2M(x)Q(x)')\frac{\partial^2\phi}{\partial x^2}+
Q(x)\frac{\partial^2\phi}{\partial t\partial x} -
M(x)Q(x)\frac{\partial^3\phi}{\partial x^3},
\end{equation}
\begin{equation}
\label{4.7}
a_{3}=2M(x)-Q(x)H(x),
\end{equation}
\begin{equation}
\label{4.8}
a_{2}=H(x)Q(x)P(x)+(2M(x)-H(x)Q(x))Q(x)'-W(x)Q(x)^2,
\end{equation}
\begin{equation}
\label{4.9}
a_1=-H(x)Q(x)P(x)'-H(x)P(x)Q(x)'-M(x)Q(x)''-V(x)Q(x)-2W(x)P(x)Q(x),
\end{equation}
\begin{equation}
\label{4.10}
a_{0}=-M(x)P(x)'' - H(x)P(x)P(x)'-W(x)P(x)^2-V(x)P(x),
\end{equation}
where primes denote differentiation with respect to $x$.
To satisfy (\ref{4.4}) it is sufficient to let $a_{21}=a_{11}=0$
and $a_3=a_2=a_1=a_0=0$.

From $a_3=0$ it follows that
\begin{equation}
\label{4.11}
Q(x)=\frac{2M(x)}{H(x)}
\end{equation}
In view of this relationship $a_{21}=0$ is satisfied in virtue of (\ref{4.2}).
To satisfy $a_{11}=0$ we add and subtract 
$M(x)'\frac{\partial^2\phi}{\partial x^2}$ and rewrite this equation in 
the form
\begin{equation}
\label{4.12}
a_{11}=-(Q(x)H(x)P(x)+2M(x)Q(x)'-Q(x)M(x)')\frac{\partial^2\phi}{\partial x^2}+
Q(x)\frac{\partial^2\phi}{\partial t\partial x} -
Q(x)\frac{\partial}{\partial x}\left( M \frac{\partial^2\phi}{\partial x^2}\right)=0.
\end{equation}
Letting 
\begin{equation}
\label{4.13}
P(x)=\frac{2M(x)H(x)'}{H(x)^2}-\frac{M(x)'}{H(x)}
\end{equation}
we infer that (\ref{4.12}) is satisfied using (\ref{4.2}).

We now use $a_2=a_1=0$ to express $W(x)$ and $V(x)$ in terms of $M(x),\,H(x)$.
Substituting the expressions for Q(x) and P(x) in $a_2=0$ we obtain 
\begin{equation}
\label{4.14}
W(x)=-\frac{H(x)M(x)'}{2M(x)}+H(x)'
\end{equation}
Similarly $a_1=0$ yields
\begin{equation}
\label{4.15}
V(x)=-\frac{M(x)H(x)''}{H(x)}
\end{equation}
With these results $a_0=0$ yield a differential equation relating
$M(x)$ and $H(x)$
\begin{equation}
\label{4.16}
\left[\frac{H(x)M(x)'}{2M(x)}-H(x)'\right]P(x)^2+
\left[\frac{M(x)H(x)''}{H(x)}-H(x)P(x)'\right]P(x)
-M(x)P(x)''=0
\end{equation}
where for brevity we did not substitute for $P(x)$.

When $M(x)=A$ where $A$ is a constant (\ref{4.16}) becomes
\begin{equation}
\label{4.17}
H(x)^2H(x)'''-5H(x)H(x)'H(x)''+4(H(x)')^3=0.
\end{equation}
To solve this equation we introduce $z(x)=\frac{1}{H(x)}$. The resulting 
equation can be written as
\begin{equation}
\label{4.18}
\frac{d}{dx}\left(\frac{1}{z(x)}\frac{d^2z(x)}{dx^2}\right)=0.
\end{equation}
Hence either 
\begin{equation}
\label{4.18a}
H(x)=\frac{1}{ax+b}
\end{equation}
where $a,b$ are constants or
\begin{equation}
\label{4.19}
H(x)=\frac{B}{\cos(\omega x + \beta)},\,\,\,\, B,\omega,\beta,\,\,\, constants,
\end{equation}
or
\begin{equation}
\label{4.20}
H(x)=Ce^{\alpha x},\,\,\,\, C,\alpha,\,\,\, constants,
\end{equation}

When $H(x)=1$ eq. (\ref{4.16}) becomes
\begin{equation}
\label{4.21}
\frac{M(x)'}{2M(x)}-M(x)'M(x)''+M(x)M(x)'''=0.
\end{equation}
Substituting $M(x)=w(x)^2$ this equation reduces to
\begin{equation}
\label{4.22}
\frac{d}{dx}\left(w(x)\frac{d^2w}{dx^2}\right) =0.
\end{equation}
Hence
$$
w(x)\frac{d^2w}{dx^2} =c
$$
where $c$ is a constant. When $c=0$ it follows that 
\begin{equation}
\label{4.23}
M(x)=(a_1x+b_1)^2
\end{equation}
where $a_1,b_1$ are constants. When $c \ne 0$ we can find an implicit 
expression $w(x)$
$$
x=C_2\pm \displaystyle\int^{w(x)}\frac{ds}{\sqrt{2c\ln(s)-C_1c}},
$$
where $C_1,\,C_2$ are integration constants.

To our best knowledge the linearization of the "modified Burger's equations" 
represented by (\ref{4.18a})-(\ref{4.20}) and (\ref{4.23}) did not
appear in the literature so far.

{\bf Example}: For $M(x)=1$, $H(x)=Ce^{\alpha x}$ we obtain from (\ref{4.14})
(\ref{4.15}) respectively that
$$
W(x)=C\alpha e^{\alpha x},\,\,\,\, V(x)=-\alpha^2.
$$
Hence the solutions of (\ref{4.1}) with these coefficients is related to the 
solutions of the Heat equation (\ref{4.2}) by the transformation
(\ref{4.3}) with 
$$
Q(x)=\frac{2e^{-\alpha x}}{C},\,\,\,\, P(x)=\frac{2\alpha e^{-\alpha x}}{C}
$$
and this fact can be verified by direct substitution. 

\setcounter{equation}{0} 
\section{Second Order Convective ODEs}

We shall say that the solutions of the equations
\begin{equation}
\label{2.1}
\psi(x)''=S(x)+ [V(x)+F(x)\psi(x)']\psi(x)+W(x)\psi(x)^2
\end{equation}
and
\begin{equation}
\label{2.2}
\phi(x)''=U(x)\phi(x)
\end{equation}
are related if we can find functions $P(x)$ and $Q(x)$ so that
\begin{equation}
\label{2.3}
\psi(x)=P(x)+Q(x)\frac{\phi(x)^{\prime}}{\phi(x)}.
\end{equation}
Furthermore we observe that (\ref{2.1}) can take the more general form
\begin{equation}
\label{2.1b}
\psi(x)'' = S(x)+ (V(x)+F(x)\psi(x)')\psi(x)+V_1(x)\psi(x)'+W(x)\psi(x)^2.
\end{equation}
In this case we can find $p(x)$ so that $V_1(x)=-2\frac{p(x)'}{p(x)}$.
Introducing $\xi(x)=p(x)\psi(x)$, (\ref{2.1b}) becomes
\begin{equation}
\label{2.1c}
\xi(x)''= p(x)S(x)+\left[V(x)+\frac{p(x)''}{p(x)}+\frac{F(x)}{p(x)}\xi(x)'\right]\xi(x)+
\left[\frac{W(x)}{p(x)}-\frac{F(x)p(x)'}{p(x)^2}\right]\xi(x)^2.
\end{equation}
which has the same form as (\ref{2.1}).

To classify those nonlinear equations (\ref{2.1}) which can be "paired"
with a linear equation of the form (\ref{2.2}) we differentiate 
(\ref{2.3}) twice and in each step replace the second order derivative 
of $\phi(x)$ by $U(x)\phi(x)$. We then use 
(\ref{2.1}) to eliminate $\psi(x)''$. As a result we find that the following 
equation must hold;
\begin{equation}
\label{2.4}
a_3(x)\left(\frac{\phi(x)'}{\phi(x)}\right)^{3}+
a_2(x)\left(\frac{\phi(x)'}{\phi(x)}\right)^{2}+
a_1(x)\frac{\phi(x)'}{\phi(x)}+a_0(x)=0
\end{equation}
where 
\begin{equation}
\label{2.5}
a_3(x)=(Q(x)F(x)+2),\,\,\,a_2(x)=Q(x)(F(x)P(x)-W(x)Q(x))-(2+F(x)Q(x))Q(x)',
\end{equation}
\begin{eqnarray}
\label{2.6}
a_1(x)&=&Q(x)''-F(x)P(x)Q(x)'-U(x)F(x)Q(x)^2 - \\ \notag
&&(2U(x)+V(x)+F(x)P(x)'+2W(x)P(x))Q(x),
\end{eqnarray}
\begin{eqnarray}
\label{2.7}
a_0(x)&=&P(x)''-W(x)P(x)^2-(F(x)Q(x)U(x)+V(x)+F(x)P(x)')P(x)+ \\ \notag
&&Q(x)U(x)'+2U(x)Q(x)'-S(x).
\end{eqnarray}
To satisfy (\ref{2.4}) it is therefore sufficient to let 
$a_i(x)=0,\,\,i=0,1,2,3$.

One can use these conditions in two ways . The first is to assume that 
a nonlinear equation (\ref{2.1}) is given and try to determine the appropriate 
$P,Q,U$ (if they exist) that relates it to (\ref{2.3}). Otherwise one may 
fix the functions $P,Q,U$ and classify those nonlinear equations of the form 
(\ref{2.1}) which are related to (\ref{2.3}) by the transformation (\ref{2.2}).
In the following we provide separate solutions for these two possibilities.

Assuming that one starts from (\ref{2.1}) i.e the functions $V(x),W(x),F(x)$
and $S(x)$ (with $F(x) \ne 0$) are given it follows from (\ref{2.4}) that
\begin{equation}
\label{2.8}
Q(x)=-\frac{2}{F(x)}.
\end{equation}
Substituting this result in (\ref{2.5}) and solving for $P(x)$ it follows that
\begin{equation}
\label{2.9}
P(x)=-\frac{2W(x)}{F(x)^2}.
\end{equation}
Using (\ref{2.8}),(\ref{2.9}) and (\ref{2.7}) we obtain a linear first order 
differential equation for $U(x)$
\begin{eqnarray}
\label{2.10}
&&U(x)'+\frac{F(x)S(x)}{2}+\frac{W(x)''-V(x)W(x)+(2W(x)-2F(x)')U(x)}{F(x)}\\ \notag
&&\frac{(2W(x)-4F(x)')W(x)'-2W(x)F(x)''}{F(x)^2}+
\frac{2W(x)(W(x)^2+3(F(x)')^2-2W(x)F(x)'}{F(x)^3}=0
\end{eqnarray}
Finally eq. (\ref{2.6}) provides (after using (\ref{2.8}) and (\ref{2.9}))
an intrinsic constraint on the functions $V(x),W(x),F(x)$and $S(x)$ which 
have to be satisfied for the relationship between (\ref{2.1}) and (\ref{2.3})
to exist.
\begin{equation}
\label{2.12}
V(x)+\frac{F(x)''-2W(x)'}{F(x)}+\frac{6W(x)F(x)'-2(F(x)')^2-4W(x)^2}{F(x)^2}=0.
\end{equation}

We observe that when $F(x)=0$ the algorithm can be implemented in
the same way by adding a term $R(x)\psi(x)^3$ to eq. (\ref{2.1}). 

For the reverse procedure where one elects the functions $U(x),P(x)$ and 
$Q(x)$ and attempts to evaluate the corresponding nonlinear equation 
(\ref{2.1})we have from (\ref{2.7})
\begin{equation}
\label{2.13}
F(x)=-\frac{2}{Q(x)},\,\,\, W(x)=-\frac{2P(x)}{Q(x)^2}.
\end{equation}
Substituting these results in (\ref{2.7}) and solving for V(x) yields
\begin{equation}
\label{2.14}
V(x)=\frac{Q(x)Q(x)''+4P(x)^2+2(P(x)Q(x))'}{Q(x)^2}.
\end{equation}
Finally from (\ref{2.8}) we derive  an expression for $S(x)$
\begin{equation}
\label{2.15}
S(x)=P(x)''+Q(x)U(x)'+2(Q(x)'+P(x))U(x)-\frac{P(x)Q(x)''}Q(x)
-\frac{2P(x)^2(Q(x)'+P(x))}{Q(x)^2}.
\end{equation}

{\bf Example}: In the differential equation
\begin{equation}
\label{2.16}
\psi(x)''= [4a^2+\psi(x)']\psi(x)+a\psi(x)^2,
\end{equation}
where  $a$ is a constant we have $F(x)=1$, $W(x)=a$, $V(x)=4a^2$ and 
$S(x)=0$. The equation satisfies the constraint (\ref{2.12}) and using
(\ref{2.8})-(\ref{2.10}) we find that
$$
Q(x)=-2,\,\,\, P(x)=-2a.
$$
From (\ref{2.10}) we find that the general solution for $U(x)$ is
$$
U(x)=Ce^{-2ax}+a^2.
$$
With this $U(x)$ the general solution of (\ref{2.2}) is 
$$
\phi(x)=C_1BesselI(1,z)+C_2BesselY(1,iz)
$$ 
where $z=\sqrt{C}\frac{e^{-ax}}{a}$ and $BesselI$, $BesselY$
are the modified Bessel functions of the first and second kind.
Letting $C_2=0$ (real solution) it is straightforward to verify that
$$
\psi(x)=-2a-2\frac{\phi(x)'}{\phi(x)}
$$
is a solution of (\ref{2.16}).

\section{Conclusions}

We demonstrated in this paper that a generalized form of Cole-Hopf 
transformation (\ref{1.3}) can be used to relate the solutions of some 
nonlinear equation with convective terms with the solutions of linear 
differential equations. The algorithm is straightforward and we applied it 
to a generalized form of Burger's equation and second order nonlinear ODEs 
with convective terms. 

\section*{References}

\begin{itemize}

\item[1] Burgers J.M. A mathematical model illustrating the theory of 
turbulence, Adv. Appl. Mech., 1948. vol. 1, pp.171-199.

\item[2] E. Hopf - The partial differential equation $u_t+uu_x=u_{xx}$,
Commun. Pure Appl. Math {\bf 3}, pp.201-230 (1950)

\item[3] J.D. Cole, On a quasi-linear parabolic equation occurring in 
aerodynamics, Quart. Appl. Math. {\bf 9}, pp.225-236 (1951)

\item[4] P. L. Sachdev-A generalized Cole-Hopf transformation for nonlinear 
parabolic and hyperbolic equations,
ZAMP {\bf 29}, No 6 (1978), pp. 963-970, DOI: 10.1007/BF01590817

\item[5] B. Gaffet-On the integration of the self-similar equations and
the meaning of the Cole-Hopf Transformation J. Math. Phys. 27, 2461 (1986)

\item[6] D. Senouf, R. Caflisch and N. Ercolani- Pole dynamics and 
oscillations for the complex Burgers equation in the small-dispersion limit
Nonlinearity {\bf 9} p. 1671 (1996) doi:10.1088/0951-7715/9/6/016

\item[7] B. Mayil Vaganan and M. Senthil Kumaran-
Exact Linearization and Invariant Solutions of the Generalized Burgers 
Equation with Linear Damping and Variable Viscosity
Studies in Applied Mathematics, 117 pp.95-108 (2006)

\item[8] B. Mayil Vaganan and T. Jeyalakshmi - Generalized Burgers Equations 
Transformable to the Burgers Equation,
Studies in Applied Mathematics, 127 pp 211-220 (2011)

\item[9] Wen-Xiu Ma- An exact solution to two-dimensional 
Korteweg-de Vries-Burgers equation,  J. Phys. A: Math. Gen. 26 L17 (1993).

\item[10] N. H. Ibragimov, M. Torrisi and R. Tracinà- 
Self-adjointness and conservation laws of a generalized Burgers equation
J. Phys. A: Math. Theor. 44 145201 doi:10.1088/1751-8113/44/14/145201(2011)

\item[11] M. Humi -A Generalized Cole-Hopf Transformation
for Nonlinear ODES, J. Phys. A: Math. Theor. 46 (2013) 325202 (14pp)

\item[12] Alain Rigal, High Order Difference Schemes for Unsteady
One-Dimensional Diffusion-Convection Problems, Journal of 
Computational Physics, Volume 114, pp. 59-76 (1994)

\item[13] S. Karaa1 and J. Zhang - High order ADI method for solving 
unsteady convection diffusion problems, Journal of Computational Physics 
Volume 198, Issue 1, pp 1-9 (2004)

\end{itemize}

\end{document}